# Deep machine learning-assisted multiphoton microscopy to reduce light exposure and expedite imaging

Stephen McAleer[1,2], Alex Fast[3], Yuntian Xue[4], Magdalene Seiler[5], William Tang[4], Mihaela Balu[3], Pierre Baldi[1,2]*, Andrew W. Browne[4,5]*

1. Department of Computer Science, University of California, Irvine, USA.
2. Institute for Genomics and Bioinformatics, University of California, Irvine, USA.
3. Beckman Laser Institute and Medical Clinic University of California, Irvine, USA.
4. Department of Biomedical Engineering, University of California, Irvine, USA.
5. Gavin Herbert Eye Institute, Department of Ophthalmology, University of California, Irvine, USA.

*Corresponding Authors

**The work described was supported by:**
Research to Prevent Blindness unrestricted grant to UC Irvine Department of Ophthalmology
UC Irvine Institute for Clinical and Translational Science -  KL2 Grant number is KL2 TR001416

**Abstract:**
**Short:** Two-photon excitation fluorescence (2PEF) allows imaging of tissue up to about one millimeter in thickness. Typically, reducing fluorescence excitation exposure reduces the quality of the image. However, using deep learning superresolution techniques, these low resolution images can be converted to high resolution images. This work explores improving human tissue imaging by applying deep learning to maximize image quality while reducing fluorescence excitation exposure. We analyze two approaches: an approach based on U-Net, and a patch-based approach on a skin dataset and an eye dataset. We find that both methods are able to successfully perform superresolution and increase image quality while reducing fluorescence excitation exposure.

**Long:** Two-photon excitation fluorescence (2PEF) allows imaging of tissue up to about one millimeter in thickness. Typically, reducing fluorescence excitation exposure reduces the quality of the image. However, using deep learning superresolution techniques, these low-resolution images can be converted to high-resolution images. This work explores improving human tissue imaging by applying deep learning to maximize image quality while reducing fluorescence excitation exposure. We analyze two methods: a method based on U-Net, and a patch-based regression method. Both methods are evaluated on a skin dataset and an eye dataset. The eye dataset includes 1200 paired high power and low power images of retinal organoids. The skin dataset contains multiple frames of each sample of human skin. High-resolution images were formed by averaging 70 frames for each sample and low-resolution images were formed by averaging the first 7 and 15 frames for each sample. The skin dataset includes 550 images for each of the resolution levels. We track two measures of performance for the two methods: mean squared error (MSE) and structural similarity index measure (SSIM). MSE measures the average squared error between predicted and true images, with a lower MSE indicating improved performance. SSIM is another method for estimating the similarity of two images with a higher score indicating two images are more similar. For the eye dataset, the patches method achieves an average MSE of 27,611 compared to 146,855 for the U-Net method, and an average SSIM of 0.636 compared to 0.607 for the U-Net method. For the skin dataset, the patches method achieves an average MSE of 3.768 compared to 4.032 for the U-Net method, and an average SSIM of 0.824 compared to 0.783 for the U-Net method. Despite better performance on image quality, the patches method is worse than the U-Net method when comparing the speed of prediction, taking 303 seconds to predict one image compared to less than one second for the U-Net method.

## 1. Introduction:

Traditional fluorescence microscopy, also known as single photon microscopy, illuminates a sample using short wavelength light to excite fluorescent molecules which fluoresce at a longer wavelength[1]. Fluorescent light is detected, and a fluorescence image reconstructed. Multiphoton microscopy, by comparison, splits the energy required for fluorescence excitation into two or three lower energy photons of light (typically infrared spectrum). Two-photon excitation fluorescence (2PEF) occurs with simultaneous absorption of two photons by a molecular fluorophore, however, 2PEF is a non-linear process proportional to the square of the instantaneous excitation light intensity [2,3]. The major benefits of 2PEF over single photon microscopy are 2-fold. 2PEF induces fluorescence excitation in a small focal volume where two photons interact simultaneously with the tissue, as opposed to single photon techniques where the fluorescence occurs along the light path length. Second, 2PEF uses infrared light to excite fluorescence, and deeper tissue penetration by infrared light than visible light allows imaging deeper into a sample.

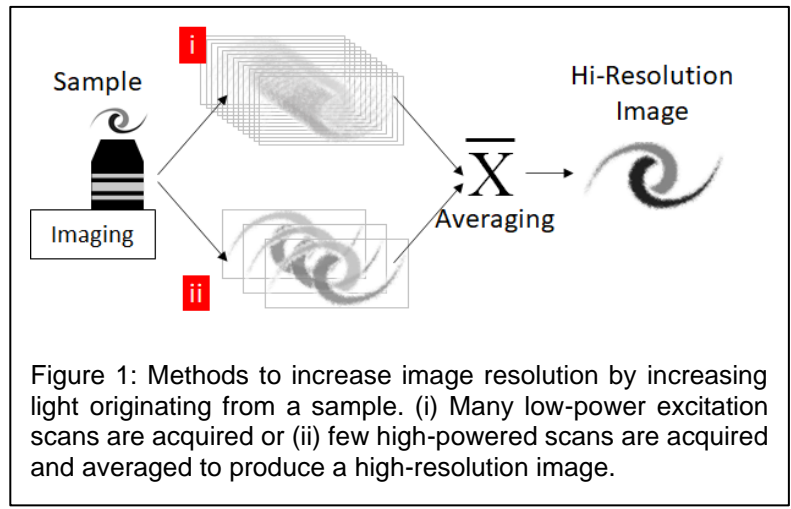

Microscopy image resolution is determined by how much focused light is detected by the imaging system. Therefore, increasing light originating from a sample, refining optical focus and optimizing light detector sensitivity each can improve image resolution. Scan averaging and higher power fluorescence excitation allow imaging systems to perceive more light from a sample to produce a higher resolution image. Acquiring many low-energy excitation images of a sample and averaging the images into a single image simulates increased light originating from the sample (**Fig. 1-i**). Alternatively, sample fluorescence is amplified by increasing excitation power and fewer scans are required to capture an image (**Fig. 1-ii**).

Figure 1: Methods to increase image resolution by increasing light originating from a sample. (i) Many low-power excitation scans are acquired or (ii) few high-powered scans are acquired and averaged to produce a high-resolution image.

Safety is a primary concern for all optical technologies used to image human tissues, especially the retina. Fortunately, all preclinical results in murine eyes demonstrate no damage from 2PEF as evaluated using visual function testing, *in* vivo structural imaging, *ex vivo* histology, and *ex vivo* biochemical analysis[4].

2PI has been comprehensively evaluated in rodents *in vitro* and *in vivo* demonstrated to be safe.[4] Briefly, 2-photon infrared light (1.2 mW laser power, imaging duration 130 seconds, 100 frames, 790nm, 20 fs pulse duration, 80MHz pulse frequency, total dose 15.6 J/cm$^2$) was compared with no light exposure and white light exposure. Structural imaging (Optical coherence tomography, confocal scanning laser ophthalmoscopy, autofluorescence), cellular electrophysiology (electroretinogram) and cellular biochemistry (quantity of rhodopsin and 11-cis retinal) were no different between eyes exposed to no light and eyes exposed to 2-photon infrared laser. However, all endpoints were significantly diminished in eyes exposed to visible white light. Schwarz et al (2016) demonstrated in macaques that infrared 2-Photon light exposure (0.5mW laser power, imaging duration 40 seconds, 900 frames, 730nm, 55 fs pulse duration, 80MHz pulse frequency, total dose 20.4 J/cm$^2$) resulted in changes in infrared autofluorescence immediately following light exposure but no detectable functional changes.[5] These changes were noted to revert to normal over a period of 22 weeks of observation. No other structural or functional alterations were detected by other imaging techniques for any of the lower dose exposures. Further evaluation in macaques demonstrated that 856J/cm$^2$ pulsed infrared 2-Photon light exposure (730nm, 55 fs pulse, 80MHz repetition frequency) resulted in changes in infrared adaptive optics findings in one class of retinal photoreceptors (s-cones, blue).[6] However, they also demonstrated that this effect was not seen with lower power laser energy doses ranging 214-489 J/cm$^2$. *D*espite the safety profile of 2PEF in preclinical models, it is essential to reduce risks of light toxicity to human tissues in every way possible. Both novel biophotonic principles and optimal data processing methods can reduce light exposure, expedite image acquisition while optimizing fluorescence image quality. This work explores improving human tissue imaging by applying deep learning to maximize image quality while reducing fluorescence excitation exposure.

Deep learning (DL) is a branch of machine learning based on artificial neural networks. Taking advantage of computing power in the form of Graphical Processing Units (GPUs) these DL approaches, in particular in the form of convolutional neural networks, are currently the method of choice in artificial intelligence and machine learning for computer vision tasks including biomedical imaging.[7-9] In image analysis, DL algorithms distinguish themselves from other approaches because they do not depend on humans to supply a list of features. DL algorithms, rather, learn relevant features directly from training data and use them for classification, regression, and other tasks. In ophthalmology, DL has been applied for instance to classify fundus photographs of diabetic retinopathy and AMD [10,11] with additional application to glaucoma screening.[12] Recently, DL has assisted fluorescence microscopy techniques by using lower resolution images to predict higher resolution images[13], and improve mobile phone microscopes[14]. Within our group, DL has also been used in many other biomedical imaging tasks, for instance to identify polyps[22], classify genetic mutations in gliomas[23], detect cardiovascular disease[24], detect spinal metastasis[25], count hair follicles[26], identify fingerprints[27], and analyze vascularization images[28]. In short, DL can be expected to help translate 2PEF microscopy from the laboratory to human application in the clinic.

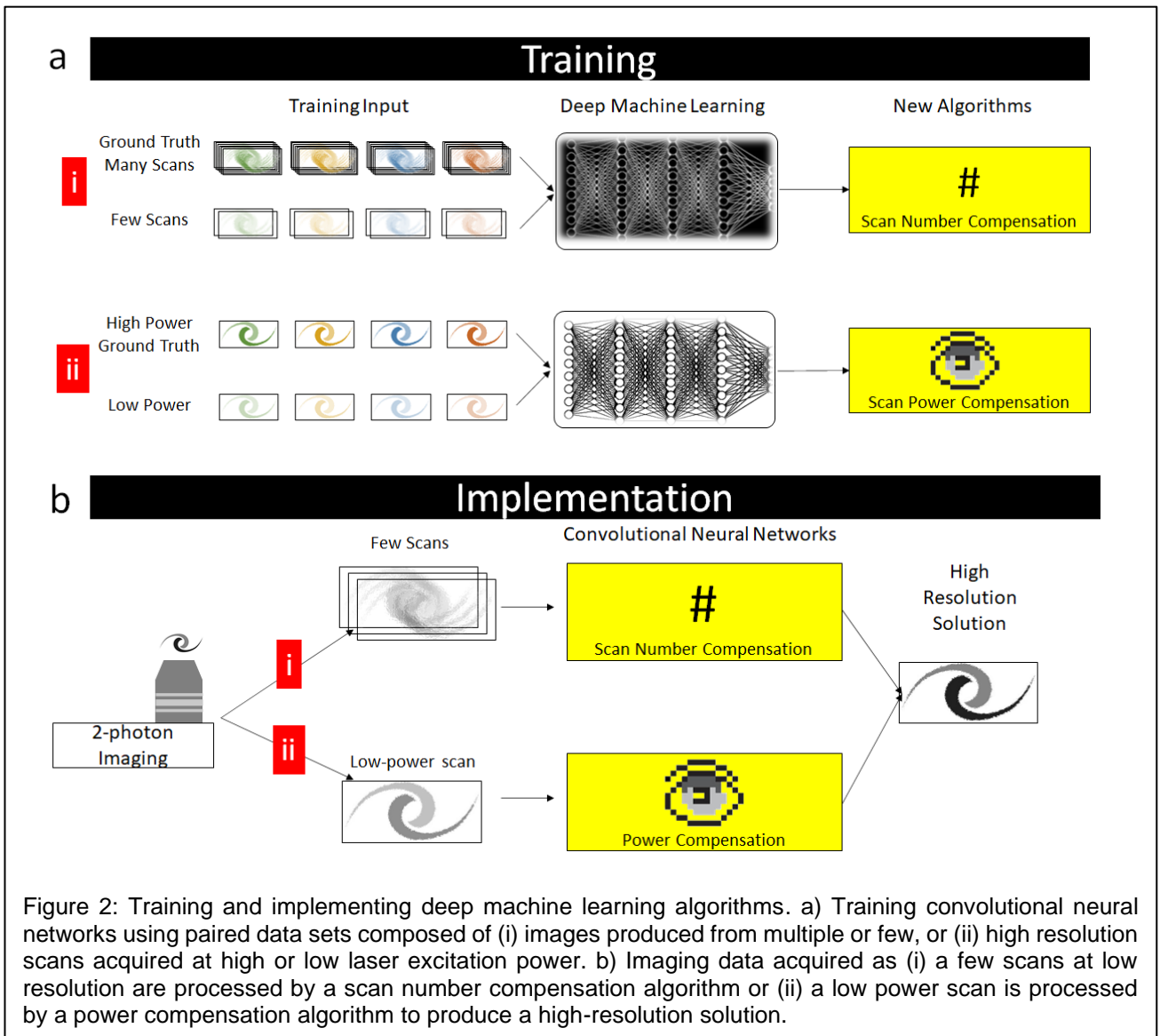

Figure 2: Training and implementing deep machine learning algorithms. a) Training convolutional neural networks using paired data sets composed of (i) images produced from multiple or few, or (ii) high resolution scans acquired at high or low laser excitation power. b) Imaging data acquired as (i) a few scans at low resolution are processed by a scan number compensation algorithm or (ii) a low power scan is processed by a power compensation algorithm to produce a high-resolution solution.

With the goal of reducing excitation light exposure and expediting imaging time we explore two strategies to assist 2PEF imaging in human stem cell derived retinal organoids[15] and human skin[15]. First, we train convolutional neural network (CNN) using paired datasets from multiple image samples acquired by averaging many scans or few scans (**Fig. 2a-i**). Second, we train a CNN using paired images acquired at low and higher scan powers (**Fig. 2a-ii**). We then finally evaluate the performance of two DL algorithms to produce high resolution images based on few input scans (**Fig. 2b-i**) or a single low power scan (**Fig. 2a-ii**).

## 2. Methods

*2.1 In vitro imaging - human retinal organoids*

A genetically modified registered embryonic line (WA01 line expressing CRX-GFP) was used for retinal organoids generation in this study. The retinal organoids in this study was produced by force aggregation method [16] and differentiated as described previously [17].

Images were taken by Zeiss LSM 510 microscope with EC Plan-Neofluar 20x/0.5 Ph2 objective. The imaging sample included 8 retinal organoids on day 61 of differentiation, each organoid was imaged in different depth and sections to obtain unique images of different cross-sections. The excitation wavelength was 740 nm, produced by Chameleon laser source, COHERENT MRU X1 (work together with cooling system – Refrigerated Recirculatory, NESLAB). The laser power values used were 300 mW (high power) and 50 mW (low power); laser pulse frequency was about 518.84 Hz as measured using a light power meter (THORLABS PM100D). Except for laser power, other imaging parameters were held constant. The scan mode was planar, 12 bits, with frame size of 512 x 512 (450 um x 450 um); pixel dwell time was 1.61 us. Laser passed through an HFT KP 650, mirror, and NFT 490 beam splitters. Each imaging region was first imaged under low power and then switch to high power. Paired images (n=300 low power, n=300 high power) were acquired with 2 separate imaging channels for nicotinamide adenine dinucleotide (390-465 nm) and flavin adenine dinucleotide (500-550 nm). A total of 1200 images were acquired to train deep learning algorithms.

*2.2 Ex vivo imaging - Human skin*

De-identified IRB exempt human skin was obtained from the UCI Dermatology Clinic and consisted of excess normal tissue discarded during wound closure procedures. The specimens were imaged fresh, immediately upon collection.

Images were taken with a home-built Fast Large Area Multiphoton Exoscope (FLAME), a device based on laser scanning multiphoton microscopy and optimized for rapid, depth-resolved imaging of large skin tissue areas with sub-micron resolution (Refs 1-3). Briefly, a frequency doubled Yb-doped 780 nm 90 fs 80 MHz amplified fiber laser (Carmel X-series, Calmar Laser, USA) was used to excite endogenous skin components such as keratin, melanin, NAD(P)H, collagen and elastin. Laser power was set to 45 mW at the focus of a 1.05 NA 25x objective lens (XLPL25XWMP, Olympus, USA) for all measurements. Emission was separated from excitation with a 705 nm shortpass dichroic mirror (FF705-Di01, Semrock, USA) and further filtered with a 620 nm shorpass filter and a 535 nm centered bandpass filter (FF01-620/SP and FF01-535/150, Semrock, USA). To detect the signal, we employed a sensitive PMT (R9880-20, Hamamatsu, Japan) in photon counting mode. A resonant scanner and galvonometric mirror pair were used to raster scan the excitation beam over the sample. Image frames were taken with 1024x1024 pixels (900x900 µm$^2$) with 88 ns dwell time per pixel and consecutively averaged. High-SNR "ground truth" images were formed by averaging 70 frames (~10 seconds) while input images for training the neural network were formed by averaging the first 7 and 15 frames (~1 and 2 seconds) of the same video stack. A total of 550 images were acquired to train deep learning algorithms from 8 excisions including pigmented and nonpigmented skin at different depths with viable epidermis.

*2.3 Deep Learning Methods*

We apply two different deep learning methods, patch-based and CARE, to the two datasets. The patch-based method involves partitioning the input into small tiles, and then training a neural network using supervised learning where the objective is to predict the high-resolution of the middle pixel. The CARE method uses U-

Net, which is a popular model that uses contractive convolutional neural networks and expansive up-convolutional neural networks for fine-grained prediction.

*2.3.1 Patch-Based Regression*

In this method, we follow the patch-based method of Ciresan et. al[21]. We first create patches of the input image. Each patch is a 40x40 pixel square around each pixel in the input image. For pixels near the edge, a 40x40 pixel square will not fit inside the image, so we pad the extra values with 0. These patches are then compiled as a set of inputs. The target corresponding to a given patch is the center pixel of the patch in the high-resolution image. The input image and target image both have two channels. We then train a neural network to predict the target pixel from the input of the patch using mean-squared-error as the loss function. We use a neural network with two convolutional layers followed by three fully connected layers. The first convolutional layer has 64 filters with a kernel size of 4x4. The second convolutional layer has 32 layers with a kernel size of 3x3. A batch normalization layer is between the two convolutional layers. The fully connected layers have sizes of 1024, 512, 32, and 2, respectively. All layers have a ReLU activation. We train the neural network with Adam optimization.

*2.3.2 U-Net*

We use CSBDeep CARE[18], a model that is based on residual version of U-Net [19]. We have a dataset including Source images and high-resolution GT images. We split the images into patches and then run the CARE U-Net model on the patches. To create the training set, we extracted 128x128 patches from each image. The depth of the U-Net is 2, the kern size is 5, the last activation is linear, and the training loss is Laplace. We train the CARE U-Net on 100 training epochs with 30 training steps per epoch, with a training learning rate of 0.0004 and a batch size of 16.

*2.4 Quantitative evaluation of deep learning output*

Quantitative comparison of images produced by the two deep learning strategies for agreement with ground truth images is conventionally performed using mean square error (MSE) and structural similarity index measure (SSIM). MSE assesses the cumulative error between two data sets (images in this context). Pixels or groups of pixels between a target and source image are compared and the average squared difference between the estimated pixel values and the actual pixels. Image degradation may occur with data compression, data lost during transmission or acquisition, or data prediction. SSIM quantifies similarity between two images to produce an image degradation metric.[20]

**3. Results**

The patch-based and CARE methods applied to two different data sets demonstrate that the patch-based regression method outperforms the U-Net method based on MSE and SSIM. However, this could be because the patch-based method was trained for longer than the CARE method. Because both methods used different architectures, training parameters, and training time, these results should not be used to infer that the patch-based method will always outperform the CARE method. Rather, these results demonstrate that both methods can achieve good performance on both datasets. Furthermore, as shown below, the patch-based method runs considerably slower and has many more parameters than the CARE method.

3.1 Retinal Organoid Data

Ground truth images were acquired using high laser power (300mW) while input images for restoration were acquired using low laser power (50mW). Among the 1200 images acquired at each power, 1169 images were used for neural net training, while 31 images were held out for testing. MSE and SSIM were determined for the held-out test set of 31 images.

## 3.2 Skin Data

High-resolution GT images were produced from the average of 70 frames, while the low-resolution sources images were produced from the average of 7 frames. Two source images were used. The first source was formed by averaging the first 7 frames and the second source was formed by averaging the first 15 frames of the same video stack. Among the 713 images acquired at both frame counts, 641 images were used for neural net training, while 72 images were held out for testing. MSE and SSIM were determined for the held-out test set of 72 images

Figure 3 demonstrates representative images restored using the two deep learning approaches and by averaging the restored images to produce an average image produced by the two deep learning approaches. Associated MSE and SSIM images are shown for each of the 3 approaches. As shown in the table below, the patches method can achieve much lower MSE and slightly higher SSIM.

Table: Quantitative output for all test images from the retinal organoid and skin data sets. Average values were determined for 31 values in the retinal organoid set and 72 values in the skin set.

| Data Set |  | Patches | CARE | Average$^{(Patches+CARE)}$ |
|---|---|---|---|---|
| Retinal Organoid | Average MSE | **27,611** | 146,855 | 55,546 |
|  | Average SSIM | **0.636** | 0.607 | 0.620 |
| Skin Samples | Average MSE | **3.768** | 4.032 | 3.803 |
|  | Average SSIM | **0.824** | 0.783 | 0.816 |

Table: Comparison of the two methods by number of parameters and time to predict.

|  | Patches | CARE |
|---|---|---|
| Number of Parameters | 20,260,162 | **923,876** |
| Time to Predict One Image (s) | 303.68 | **0.89** |

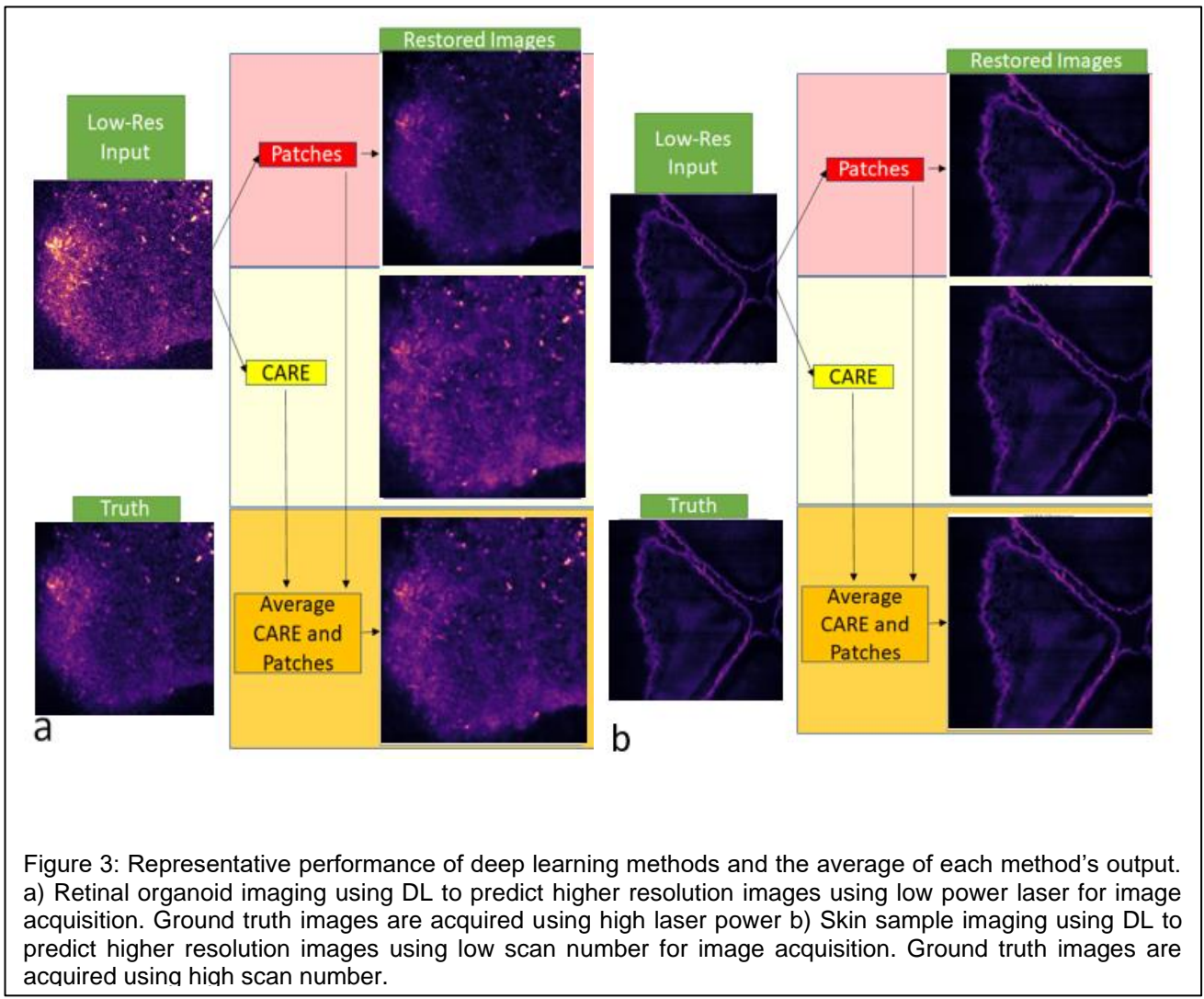

Figure 3: Representative performance of deep learning methods and the average of each method's output. a) Retinal organoid imaging using DL to predict higher resolution images using low power laser for image acquisition. Ground truth images are acquired using high laser power b) Skin sample imaging using DL to predict higher resolution images using low scan number for image acquisition. Ground truth images are acquired using high scan number.

## 4. Discussion

One of the primary concerns with the application of advanced laser imaging to humans is safety. The retina is the tissue most sensitive to light. Therefore, minimizing the light exposure needed to image the retina can aid imaging all tissues in humans in vivo. In general, image quality, in optical imaging, is improved by increasing the light captured. In multiphoton microscopy, fluorescence is excited by a ultrafast femtosecond laser and the emitted light is detected. We sought to determine if multiphoton excitation of intrinsic fluorophors to image human skin and retinal organoid tissue can be enhanced using deep learning methodology.

In this paper we present two deep learning methods for improving the image quality of microscopy images acquired using 2-photon excited fluorescence. The first method is based on U-Net and the second method is a patch-based regression model. We evaluate these methods on two datasets. The first dataset is composed of human retinal organoids and the second is composed of human skin. We find that both methods are able to achieve good performance on both of the datasets. In other words, deep learning is able to reconstruct high resolution images from lower resolution images acquired using lower excitation laser power or fewer excitation light scans. We find that with more training time, the patch-based regression approach is able to achieve higher image quality than the U-Net, but is much slower at predicting new images.

In conclusion, machine learning has been a valuable tool to enhance multiple microscopy techniques. We demonstrated that deep learning will be a valuable approach to retaining high resolution imaging using multiphoton microscopy of even retinal tissues, while minimizing light exposure needed to acquire an image.